\documentclass[colorlinks=true, citecolor=blue, linkcolor=red, urlcolor=black,
reprint,
superscriptaddress,
nofootinbib,
amsmath,amssymb,
aps,
]{revtex4-2}
\usepackage{amsmath}
\usepackage{color}
\usepackage{xcolor}
\usepackage[table]{xcolor}
\usepackage{graphicx}
\usepackage{multirow}
\usepackage{booktabs}
\usepackage{array}
\usepackage{makecell}
\usepackage{orcidlink}
\usepackage{caption}
\usepackage{subcaption}
\usepackage[T1]{fontenc}
\usepackage{dcolumn}
\usepackage{bm}
\usepackage{stackengine}

\usepackage{array}
\usepackage{graphicx,pstricks,listings,stackengine}
\usepackage{tikz}
\usetikzlibrary{shapes.geometric, arrows.meta, positioning}
\usetikzlibrary{positioning,fit,backgrounds}
\usepackage{enumitem,amssymb}
\usepackage[normalem]{ulem}
\makeatletter
\count@=`A \advance\count@\m@ne
\@whilenum\count@<`Z\do{%
  \advance\count@\@ne
  \begingroup\uccode`a=\count@
  \uppercase{\endgroup\DeclareMathSymbol{a}}{\mathalpha}{operators}{\count@}%
}
\makeatother

\makeatletter
\count@=`a \advance\count@\m@ne
\@whilenum\count@<`z\do{%
  \advance\count@\@ne
  \begingroup\uccode`a=\count@
  \uppercase{\endgroup\DeclareMathSymbol{a}}{\mathalpha}{operators}{\count@}%
}
\makeatother

\begin{document}
\preprint{nsbh-eos}

\title{Constraining Equation of State of neutron star using neutron star-black hole mergers}
\author{Vasudev Dubey\, \orcidlink{0009-0009-1078-4949}}
\email{vasudev.dubey@iucaa.in}
\affiliation{Department of Physics, Indian Institute of Technology Bombay, Mumbai, Maharashtra, 400076, India}
\affiliation{Inter-University Centre for Astronomy $\&$ Astrophysics, Pune, Maharashtra, 411007, India}
\author{Rahul Kashyap\, \orcidlink{0000-0002-5700-282X}}
\affiliation{Department of Physics, Indian Institute of Technology Bombay, Mumbai, Maharashtra, 400076, India}
\author{Yugesh Bhoge\, \orcidlink{0009-0000-8556-8671}}
\affiliation{Department of Physics, Indian Institute of Technology Bombay, Mumbai, Maharashtra, 400076, India}

\date{\today}

\begin{abstract}
In this work, we study tidal effects that can be observed in gravitational wave transient from neutron star-black hole (NSBH) mergers. It has been well known that neutron star's tidal deformation has imprint of equation of state and hence composition of neutron star matter. Compared to binary neutron star mergers, where we observe a combined effect of two neutron stars by measuring the parameter known as lambda tilde. In neutron star black hole mergers, we measure the tidal deformability of the neutron star component clearly. In this work, we study the feasibility of measuring the bare tidal deformability of a neutron star with planned gravitational detectors in the near and far future. We use the parameter estimation computation framework, Bilby is used to perform Bayesian inference for hundreds of NSBH mergers in current and future GW detector networks and obtain the posterior probability distribution functions of binary parameters. We find that we need at least 20 events to clearly distinguish the equation state of neutron stars in future detectors, including the Einstein Telescope and Cosmic Explorer.
\end{abstract}

\maketitle

\section{\label{sec:intro}Introduction}
The internal structure and composition of neutron stars is still one of the central open problems in astrophysics. When two compact objects (e.g., neutron stars) inspiral, their mutual gravitational attraction deforms each body. This tidal deformation produces additional multipole moments that radiate in GWs, modifying the amplitude and phase of the waveform. In the post-Newtonian (PN) approximations, these effects are parameterized using tidal Love numbers ($k_l$) or, equivalently tidal deformabilities ($\Lambda_i$, $\tilde\Lambda$) \cite{dan2026_modeling_phase_transitions}. Although numerous equations of state (EOS) have been proposed to model neutron star structure, gravitational wave (GW) observations may provide a more effective way to constrain these properties \cite{Clarke2023-bc}. Mergers involving neutron stars, such as binary neutron star (BNS) or neutron star-black hole (NSBH) systems, offer critical insights into tidal properties under those extreme-density conditions. 

To date, relatively few BNS (GW170817  \cite{Abbott_2018_GW170817}, GW190425 \cite{Abbott_2020_GW190425}) and NSBH (GW200105, GW200115 \cite{Abbott_2021_NSBH_discovery} and GW230529 \cite{Abac_2024_GW230529}) mergers have been observed compared to BBH events, primarily due to the sensitivity limitations of current ground-based detectors. This observational gap limits our understanding of the structure and dynamics of neutron stars in regions of extreme density and strong gravity, where a different equation of state predicts different properties \cite{Zhu2022-mc,Boersma2022-wo}. 
 
With the advent of next generation detector configurations, such as O5, A$^\sharp$ (As, from now onwards), Einstein Telescope and Cosmic Explorer (ECC)\cite{O5_3G_StandardReferences, MM_of_NSBH_in_O4/5_2024, O5detections_Muccillo_2026, 3G_detectors_GW__TintoDhurandharRaj_2026, 3G_GWdetector_article_2022}, which will offer enhanced sensitivity over current detectors, we expect a significant increase in the detection rate of neutron star containing binaries \cite{Gupta_2024_Characterizing_GW_detectors}. For instance, recent studies indicate that a network of Cosmic Explorer and Einstein Telescope could detect over $3\times10^{5}$ binary neutron star mergers annually. By analysing just the 75 loudest events, studies have shown it is possible to constrain the neutron star radius to within 200 meters, representing a tenfold improvement over current LIGO and NICER results.\cite{Third_gen_detectors}. Future GW detectors such as Einstein Telescope is going to measure parameters of these sources with much greater precision providing us clues to their dynamical formation, EOS, cosmological expansion and nature of Gamma-ray bursts \citep{Abac2026-ET}.

\textcite{Sarin2024-ea} provide joint constraints on mass distribution as well as the EOS using non-parametric method and provide about 1.6\% and 13\% precision for radius and mass of a 1.4 M$_\odot$ NS, respectively. In principle, it is possible to constrain neutron star radius by measuring the disruption in GW by observing the tidal disruption frequency. However, it has been found not to be informative enough till now to constrain the radius, hence constrain the equation of state \citep{Clarke2023-bc}. \citet{170817_190425_EOS_measurements_Vivanco_2020} constrained the NS EOS by combining gravitational wave observations GW170817 and GW190425 through random-forest-regressed marginalized likelihoods, obtaining $R_{1.4}=11.6^{+1.6}_{-0.9} \ km$ at $90\%$ credibility. Complimentarily, several NS radius constraints based on electromagnetic (EM) counterpart have been proposed. For example, \citet{EOS_constraint_gw170817_review_raithel2019} found the $90\%$ highest posterior density interval of $9.8<R<13.2 \ km$ from the posterior of $\tilde\Lambda$ of $GW170817$. In addition, combining gravitational wave data with the radiative transfer models of the kilonova lightcurve and GRB jets led to proposed common radius constraints of $[11.1,13.4] \ km$. Combining the GW signal and EM counterpart constraints from GW170817 with chiral-effective-field-theory-informed equations of state, a new and significantly tighter inferred radius of neutron star is constrained as $R_{1.4}=11.0^{+0.9}_{-0.6} \ km$\cite{GW_EM_XEFT_NSradii_constraint_Capano_2020}. \citet{Fragione2021-yg} diagnostic works most efficiently when the aligned component of the BH spin is maximum, because higher BH spin can increase the tidal disruption radius enough to allow a neutron star to be torn apart outside the event horizon, enabling the debris to form an accretion disk. This will require a distinction of EM counterparts between BNS and NSBH systems \citep{Gupta2026-ng}. With an assumption of an observed associated short gamma-ray bursts (SGRB) \citep{Ascenzi2019-ae}, the NS radius can be constrained below 20\% which rely crucially on effectiveness of the follow-up in the coming era by high-energy missions. In contrast our present work rely exclusively on GW signal alone do not assume any information about disruption of NS and hence existence of EM counterparts which could be as low as 14\% of total NSBH events \citep{Biscoveanu2022-uo}.

A fundamental, distinguishable feature between neutron stars and black holes, beyond the unresolved issue of the lower mass gap\cite{Ye2022-dd}, is the non-zero tidal deformability of neutron stars. Unlike black holes, which have zero tidal deformability\citep{Poisson_2009, Chiara_2025}, neutron stars experience tidal distortion under strong gravitational fields. This property leaves measurable imprints on the GW signals. As a result, NSBH systems offer a cleaner environment for inferring the tidal properties of neutron stars, due to the asymmetry in tidal deformabilities compared to BNS systems, where mutual tidal interactions and symmetry introduce degeneracies\cite{Kyutoku2021-mh, Clarke2023-bc}. In agreement with \textcite{Biscoveanu2022-uo}, we do find that constraining EOS using BNS events are more effective than NSBH. However, our method can find correct EOS with slightly large number of NSBH events. If an NSBH merger is misidentified as a BNS system, the inferred tidal properties will appear as outliers when compared to the rest of the neutron star population.

Defining the mass ratio as $q \equiv m_{NS}/m_{BH} = m_2/m_1$, the effective tidal deformability can be expressed as:

\begin{equation}
\begin{aligned}
\tilde{\Lambda} &= \frac{16}{13} \frac{(m_1+12m_2)m_1^4\Lambda_1 + (m_2+12m_1)m_2^4\Lambda_2}{(m_1+m_2)^5} \\
\tilde\Lambda &= \frac{16}{13} \frac{(12+q)q^4\Lambda_2}{(1+q)^5}
\end{aligned}
\label{eq:lambda_eff}
\end{equation}\cite{Sieniawska_2019_Tilde_Lambda_expression}

While for the asymmetric NSBH binaries, the neutron star, in general, swallowed by the companion black hole without the sign of tidal disruption; for (near-)symmetric binaries ($q \rightarrow 1$), the tidal signature of neutron star is easy to capture. From eq-\ref{eq:lambda_eff}, for $q\rightarrow1$, $\tilde\Lambda \rightarrow \Lambda_2/2$, while for $q \rightarrow 0$, $\tilde\Lambda \rightarrow 0$, regardless of the $\Lambda_2$. So, for more symmetric systems, the tidal deformability is more detectable compared to the asymmetric case.

The primary objective of this study is to constrain the equation of state (EOS) of neutron stars by leveraging their tidal properties as key observables. Specifically, we aim to recover the injected physical parameters of the system with high accuracy. To achieve this, we employ a Bayesian inference framework that effectively incorporates prior astrophysical knowledge about the binary components. In our analysis, we utilize \texttt{IMRPhenomNSBH} gravitational waveform model to explore systematic dependencies. By varying the neutron star EOS across simulations, we assess the corresponding imprints on the gravitational wave signals and quantify the sensitivity of waveform morphology to the tidal response of neutron stars.

This study is a direct continuation of the work by \textcite{Kashyap_2025_BNS_Bayesian_optimization}, who performed the Bayesian Evidence calculation fOr Model Selection (BEOMS) framework to systematically compare EOS models using posterior distributions derived from EOS-agnostic Bayesian inference. Their analysis demonstrated that Bayesian model selection achieves peak efficiency in the two-dimensional subspace of component mass and tidal deformability $(m - \Lambda)$, as this focused dimensionality effectively avoids the `Occam’s penalty' inherent in higher dimensional calculations while requiring fewer events to distinguish between models with high confidence.

Recent investigations by \textcite{Cho_2022} have examined the broader EOS inference problem using a population of NSBH events. This study measures the accuracy with which neutron-star tidal deformability can be measured from NSBH observations. Using the Fisher matrix analysis, with 4 different waveforms and $10^3$ sources sampled randomly, the study concluded that with the 2G detectors, the accuracy $\sigma_{\Lambda_{NS}} \sim 130$, while with the third generation configuration, it is just a factor of $4$. Building on such efforts, the present work investigates EOS discrimination through Bayesian model selection using cumulative evidence from multiple NSBH detections.

This work is presented here as follows: In section-\ref{sec:method}, we describe the Bayesian inference framework employed in this study, with a particular focus on the BILBY and evidence calculation strategy using BEOMS. We discuss different waveform approximants used in our work. In section-\ref{sec:results}, we present the key results of the work, including the improvement in the parameter estimation pipeline and the resulting enhancements in efficiency. In section-\ref{sec:discussion}, we discuss the implications of the findings of this work, summarise the findings of the study and discuss their broader implications for the study of neutron stars' equation of state.

\section{Methodology}\label{sec:method}

\subsection{Bayesian Inference, Parameter Estimation using BILBY and BEOMS} \label{sec:BILBY}
We employ a Bayesian Inference framework to estimate the physical parameters of the NSBH system and to constrain the equation of state of neutron stars. In this framework, we compute the posterior probability distribution of parameters $\theta$, given gravitational wave data $d$, using Bayes' Theorem,

\begin{equation} 
    p(\theta|d) = \frac{\mathcal{L}(d|\theta) \pi(\theta)}{\mathcal{Z}}
    \label{eq:posterior}
\end{equation}

Here, $\mathcal{L}(d|\theta)$ is the likelihood function of GW data $d$ for given injected parameters $\theta$; $\pi(\theta)$ is the prior distribution of the parameters, and $\mathcal{Z}$ is a normalization factor, called "evidence", which plays an important role in model selection \cite{Thrane_2019}. In gravitational-wave analysis, the likelihood is modelled assuming Gaussian noise, and the signal template $h(\theta)$ is compared to the data using the noise-weighted inner product.

We define priors for both intrinsic as well as extrinsic parameters focusing on chirp mass ($\mathcal{M}$), mass ratio ($q$) and effective tidal deformability ($\tilde\Lambda$) for EOS inference. To implement this framework in practice, we use \texttt{Bilby} \cite{bilby}. The analysis begins by injecting simulated gravitational-wave signals $h(t)$ into synthetic strain data $n(t)$ using injected source parameters. After the injection, the waveform is generated using the waveform approximants. The choice depends on the system and purpose of analysis. For our purpose, we are primarily using \textbf{\texttt{IMRPhenomNSBH}} \cite{IMRPhenomNSBH_file_reference} to model the tidal effects in NSBH systems. This is a frequency-domain phenomenological model built specifically for NSBH binaries. This is based upon the BBH phenomenological baseline in the IMRPhenom family, and consistent with the parameter conversion we have used. We also conducted parameter estimation using \texttt{SEOBNRv4\char`_ROM\char`_NRTidalv2\char`_NSBH} \cite{SEOBNR_2020}. This is a reduced-order-model (ROM) version of the SEOBNRv4 effective-one-body (EOB) waveform, with NRTidalv2 tidal corrections added for NSBH systems. We compared our results against our standard waveform model and find consistent results.

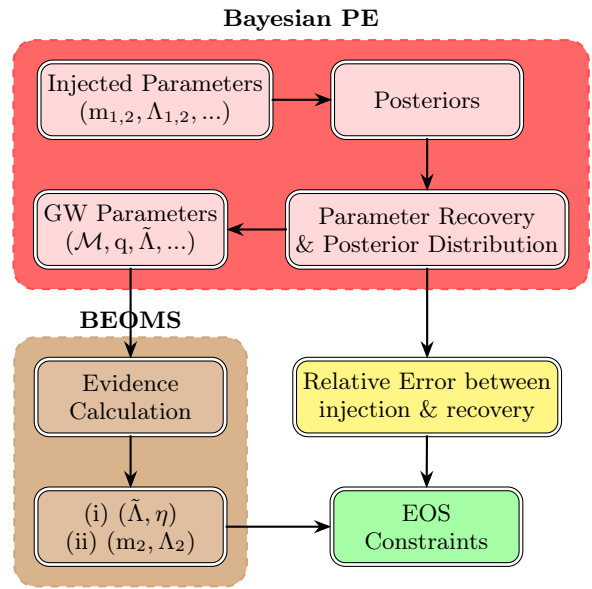
\begin{figure}[h]
\centering

\begin{tikzpicture}[
  node distance=0.7cm and 0.8cm,
  every node/.style={font=\small},
  box/.style={
    draw,
    rounded corners=5pt,
    minimum width=2.5cm,
    minimum height=1cm,
    align=center,
    fill=#1
  },
  arrow/.style={-Stealth, thick},
]

\node[box=pink!60, double, double distance=1pt]
(input) {Injected Parameters\\
$(m_{1,2}, \Lambda_{1,2},...)$};

\node[box=pink!60, double, double distance=1pt,
      right=of input]
(posterior) {Posteriors};

\node[box=pink!60, double, double distance=1pt,
      below=of posterior]
(recovery) {Parameter Recovery\\
\& Posterior Distribution};

\node[box=pink!60, double, double distance=1pt,
      left=of recovery]
(gwparams) {GW Parameters\\
$(\mathcal{M}, q, \tilde{\Lambda},...)$};

\node[box=brown!50, double, double distance=1pt,
      below=1.2cm of gwparams]
(evidence) {Evidence\\
Calculation};

\node[box=brown!50, double, double distance=1pt,
      below=of evidence]
(para_space) {(i) $(\tilde{\Lambda},\eta)$\\
(ii) $(m_2,\Lambda_2)$};

\node[box=green!35, double, double distance=1pt,
      right=1.4cm of para_space]
(eos) {EOS\\
Constraints};

\node[box=yellow!60, double, double distance=1pt,
      below=1.2cm of recovery]
(lambda_m) {Relative Error between\\
injection $\&$ recovery};

\begin{scope}[on background layer]

\node[
    draw=red!70,
    dashed,
    thick,
    rounded corners=8pt,
    fill=red!60,
    inner sep=8pt,
    fit=(input)(posterior)(recovery)(gwparams),
    label={[font=\bfseries]above:Bayesian PE}
] {};

\node[
    draw=brown!70,
    dashed,
    thick,
    rounded corners=8pt,
    fill=brown!60,
    inner sep=8pt,
    fit=(evidence)(para_space),
    label={[font=\bfseries]above:BEOMS}
] {};

\end{scope}

\draw[arrow] (input) -- (posterior);
\draw[arrow] (posterior) -- (recovery);
\draw[arrow] (recovery) -- (gwparams);
\draw[arrow] (recovery) -- (lambda_m);
\draw[arrow] (lambda_m) -- (eos);
\draw[arrow] (gwparams) -- (evidence);
\draw[arrow] (evidence) -- (para_space);
\draw[arrow] (para_space) -- (eos);

\end{tikzpicture}
\caption{Here is illustrated the data analysis pipeline adopted in this work. The workflow begins with injection of parameters for each NSBH, particularly the neutron star mass $m_{NS}$ and tidal deformability $\Lambda_{NS}$ corresponding to the injected EOS. The standard Bayesian parameter estimation is performed to obtain the posterior distributions of the source parameters. These posteriors are used to assess parameter recovery and derive better-constrained GW quantities like $\mathcal{M}$, $\mathcal{\eta}$ and $\tilde\Lambda$.After that, the Bayesian evidence ($\mathcal{Z}$) and Bayes' Factor ($\mathcal{BF}$) are evaluated in two different 2D parameter subspaces: (i) $\mathcal(\tilde\Lambda, \eta)$, (ii) $\mathcal(m_2,\Lambda_2)$. Finally, the cumulative log-evidence and the $\mathcal{BF}$ from multiple NSBH binaries are used to constrain the equation of state of neutron stars.}
\end{figure}

The priors are chosen to balance recovery accuracy and computational cost. For instance, the chirp mass $\mathcal{M}$ is sampled within a narrow range around the injected value to improve convergence. We evaluate the likelihood using the \texttt{RelativeBinningGravitationalWaveTransient} likelihood class in Bilby, assuming Gaussian noise, and the strain data is generated using the power spectral densities of corresponding detector configurations. Posterior sampling is performed using the nested sampling, which also estimates the Bayesian evidence $\mathcal{Z}$ for model comparison.

\subsection{Parameter Estimation Runs}
For this study, we employed the \texttt{IMRPhenomNSBH}\cite{IMRPhenomNSBH_file_reference} waveform model, which includes the tidal effects in the system. The \texttt{SEOBNRv4\char`_ROM\char`_NRTidalv2\char`_NSBH}\cite{SEOBNR_2020} was also incorporated into our framework; however, only preliminary tests were performed, and the full analysis pipeline was not executed with this waveform.

We sampled around the chirp mass $\mathcal{M}$ and mass ratio $q$ uniformly using \texttt{UniformInComponentChirpMass} and \texttt{UniformInComponentMassRatio}. We can recover the chirp mass very well relative to the component masses, since the phase evolution of gravitational waves is directly correlated with it. However, as we injected $m_1$ (BH mass) and $m_2$ (NS mass), these prior distributions are useful when the chirp mass and mass ratio are sampled, while the prior in component masses is uniform. The chirp mass has a different prior range corresponding to different detector configurations. For $O5$ and $As$, we sampled the chirp mass around $\pm 5\%$ of the injected value, while for the case of $ECC$, we used the sampling range around $\pm 0.05\%$ of the injected value. Our study utilizes the relative-binning technique to perform parameter estimation and to improve the likelihood calculation speed. However, this likelihood is highly sensitive to the chirp mass; a sampling range that is appropriate for a lower-sensitivity O5 or As network might not be suitable for a highly sensitive ECC configuration. Our decision to adopt a narrower prior range for the more sensitive detector can be justified by the fact that low-frequency gravitational-waveform data allow for very precise inference of the chirp mass. The secondary mass has an additional prior constraint with $[1,3]M_\odot$ range, considering that neutron star mass can not exceed the $3 M_\odot$ boundary. 

For the population of the simulation, we assumed that the components have aligned spins. While the neutron star is considered to be nearly non-rotating, and hence the $\chi_2$ is uniformly distributed in the range $[0, 0.1]$, the black hole's spins ($\chi_1$) are picked randomly from the uniform distribution $[0, 0.99]$. The sources are uniformly sampled from $10 \ Mpc$ up to a distance $500 \ Mpc$ in the comoving volume. The luminosity distance is uniformly distributed from redshift $z=0.001$ to $z=0.1$ using the Madau-Dickinson star formation rate for bns merger rates\footnote{Although \citet{2025_Harry_Hoy} placed an upper limit on the merger rate as $R_{90} = 79 \ Gpc^{-3}yr^{-1}$ inside $90\%$ confidence interval.}. The sky locations $(ra, \ dec)$, inclination $(\theta_{jn})$ and polarization angle $(\psi)$ are sampled uniformly over the sphere. The phase at the time of merger $(\phi)$ is considered 0\footnote{We updated this to $\mathcal{U}[0,2\pi]$ when worked with \texttt{SEOBNRv4\char`_ROM\char`_NRTidalv2\char`_NSBH}}. Lastly, we sampled the tidal effects with the help of $\tilde\Lambda$ and $\delta\tilde\Lambda$ parameters with the component tidal deformabilities $\Lambda_1$ and $\Lambda_2$ extracted from $(\tilde\Lambda, \delta\tilde\Lambda)$ priors. Both of these effective tidal parameters are constrained in the ranges $[0, 2\tilde\Lambda_{inj}]$ and $[-2000,2000]$, respectively. See TABLE-\ref{tab:priors_table} for prior distribution summary.

We employed the relative-binning to efficiently evaluate the likelihood of the simulated transients. We also compared two sampling algorithms, \texttt{dynesty}\cite{dynesty} and \texttt{nessai}\cite{nessai, Williams:2021qyt, Williams:2023ppp}, and found that \texttt{nessai} provides a substantially more computationally efficient sampling strategy while yeilding statistically consistent results (See \ref{sec:cost_effective} for more details). Owing to its computational efficiency and comparable inference performance, all results presented in this work are based on the analyses performed with the \texttt{nessai} sampler.

\begin{table}
    \centering
    \begin{tabular}{|c|c|}
    \hline
        \textbf{Parameters} & \textbf{Priors}\\
        \hline
        $\mathcal{M}$ (O5/As) & \texttt{UniformInComponent}($\pm 5\% \mathcal{M}_{inj})$  $M_\odot$\\
        $\mathcal{M}$ (ECC) & \texttt{UniformInComponent}($\pm 0.05\% \mathcal{M}_{inj})$  $M_\odot$\\
        $q$ & \texttt{UniformInComponent}(0.125,1)\\
        $m_2$ & Constraint[1,3] $M_\odot$\\
        \hline
        $a_1$ & $\mathcal{U}[0,0.99]$\\
        $a_2$ & $\mathcal{U}[0,0.1]$\\
        \hline
        $d_L^{com}$ & $\mathcal{U}[10,500]$Mpc\\
        RA, Dec, $\theta_{jn}$, $\psi$ & Uniform over the sphere\\
        $\phi$ & 0\\
        \hline
        $\tilde\Lambda$ & $[0,2\tilde\Lambda_{inj}]$\\
        $\delta\tilde\Lambda$ & $[-2000,2000]$\\
        \hline
    \end{tabular}
    \caption{Prior distributions adopted for the Bayesian parameter estimation of the simulated NSBH signals.}
    \label{tab:priors_table}
\end{table}

\subsection{EVIDENCE CALCULATIONS FOR NSBH MERGERS}\label{sec:evidence_calc_methods}
In this section, we will calculate the method for finding evidence for different EOS. We can compare the evidence of all selected model EOS with each other and will calculate the Bayes Factor (BF). Posterior density function for NSBH mergers can be written using Bayes' theorem as:

\begin{equation}
    p(\Theta|d_n,H_{GR},I) = \frac{\mathcal{L}(d_n|\Theta,H_{GR},I) \mathcal{\pi}(\Theta|H_{GR},I)}{\mathcal{Z}_{GR}}
    \label{eq:BAYES_poseterior}
\end{equation} 

 $\Theta$ are all the intrinsic as well as the extrinsic binary parameters; $d_n$ is the recovered data corresponding to each parameter; $H_{GR}$ is the hypothesis that the general relativity is true, and then $\mathcal{Z}_{GR}$ is the Bayesian evidence assuming $H_{GR}$ is correct. Numerically, $\mathcal{Z}_{GR}$ is the integral of the numerator over the parameter space. $\mathcal{L}(...)$ is the likelihood of the parameters recovered for the injected $\Theta$ parameters, and $\pi(...)$ is the prior distributions for the parameters. 
 
 Considering $H_k : \Lambda_k(m_{NS})$ as the hypothesis for particular EOS, and following the discussion from equation-6 of \textcite{Kashyap_2025_BNS_Bayesian_optimization}, we can calculate the evidence in two different subspaces, namely - ($m_{NS}, \Lambda_{NS}$) and ($\tilde{\Lambda}, \eta$).

\subsubsection{Evidence calculation in ($m_{NS}, \Lambda_{NS}$) space}\label{sec:Evidence-B}
For an EOS model, $k$ (i.e. hypothesis, $H_k$), the evidence for the ($m_{NS}-\Lambda_{NS}$) space is 

\begin{equation}
    \mathcal{Z}_k =
    \int dm_{NS}
    \int d\Lambda_{NS}\,
    \delta\left[
        \Lambda_{NS}-\Lambda_k(m_{NS})
    \right]
    p(m_{NS},\Lambda_{NS}|d_n)
\label{eq:mlam2_evidence}
\end{equation}
where the $p(m_{NS},\Lambda_{NS}|d_n)$ is the posterior of the NS parameters for a given NSBH events, represented by data, $d_n$. The posterior of the NS is obtained by marginalizing over the mass and its tidal deformability.

\subsubsection{Evidence calculation in ($\tilde{\Lambda}, \eta$) space}\label{sec:Evidence-A}
Instead of taking secondary parameters, we can similarly calculate evidence in the subspace of effective tidal deformability and symmetric mass ratio as follows:

\begin{equation}
    \mathcal{Z}_k =
    \int d\eta
    \int d\tilde{\Lambda}\,
    \delta\left[
        \tilde{\Lambda}
        - \tilde{\Lambda}_k(\eta;\bar{\mathcal{M}})
    \right]
    p(\eta,\tilde{\Lambda}|d_n)
\label{eq:etalam_evidence}
\end{equation}
where 
$p(\eta,\tilde\Lambda|d_n)$  
is the posterior obtained by marginalizing over all parameters except $\tilde{\Lambda}$ and $\eta$ including the mean chirp mass ($\bar{\mathcal{M}}$) which is known with much higher accuracy than $\tilde{\Lambda}$ and $\eta$. The effective tidal deformability, $\tilde{\Lambda}$ in the Dirac delta function is evaluated for an EOS model, $k$ and mass of the NS ($m_{NS}$) denoted in \ref{eq:etalam_evidence} as $\tilde{\Lambda}_k(\eta; \bar{\mathcal{M}}) = \tilde\Lambda_k(\Lambda_{NS}(\bar{\mathcal{M}}, \eta), \Lambda_{BH}=0, \eta)$. 

\subsection{CUMULATIVE EVIDENCE FROM MULTIPLE EVENTS \& BAYES FACTOR}
For the model selection calculation, we can find the odds ratio, $[O^i_j]_n$, given by the ratio of the evidence values for the $\textit{i}^{th}$ and $\textit{j}^{th}$ model EOS and event, $n$.

If we consider equal prior distributions for all EOS, the combined odd ratio, $log \mathcal{O}_j^i$ can be written as the difference between $log\mathcal{Z}$ of two EOSs.

But as the evidence depends on likelihood and prior, we will have to consider a correction factor. When we compute "per-event evidence" $\mathcal{Z}_k$, for a hypothesis $H_k$, the integrals are restricted to the EOS-specified allowed prior region. So, different EOS hypotheses implicitly carry unique and different upper bounds on the neutron-star mass ($m_{NS}^{max}$), and hence the prior volumes will not be identical across EOS models. As a result, this introduces a systematic bias when comparing evidence and calculating Bayes' Factor for a population. To correct this, we apply an \textit{EOS-specific prior correction factor}, ${f}$ for comparing two different EOS $i$ and $j$:

\begin{equation}
    {f} = \frac{M_{max,j} - M_{min}}{M_{max,i} - M_{min}}
    \label{eq:Corr_Factor_in_OddRatio}
\end{equation}

For the cumulative odd ratio over a population of N events, this correction factor will be exponentiated with a number $k$, which depends on the dimensions of the parameter subspace and the total number of events. For the BNS case in the $(\tilde\Lambda,\eta)$ subspace, both NS contribute to the evidence calculation. As each event has two masses that are constrained by the EOS maximum allowed mass, for $N$-events, there are $2N$ NS-level contributions. Since we converted evidences with different priors to a common prior convention to pick one reference prior, that choice introduces one global normalization that cancels out one factor. Consequently, the effective exponent is  $k=2N-1$ for the BNS-$(\tilde\Lambda,\eta)$ case. 

However, NSBH systems have only one NS, and therefore contributes one EOS-dependent  factor. After accounting for the same global normalization, the corresponding exponent is $(N-1)$. In the $(m_i,\Lambda_i)$ subspace, the evidence is calculated for individual NS separately, hence the same counting applies, giving $(N-1)$ for both BNS and NSBH systems. The value of $k$ in each parameter subspace for NSBH and BNS is summarized in  TABLE-\ref{tab:k_values}.

\begin{table}[h]
\centering
\begin{tabular}{|@{}c | c c@{}|}
\toprule
\textbf{Parameter Spaces} & \textbf{BNS} & \textbf{NSBH} \\
\hline
$(\tilde{\Lambda},\eta)$ & 2N–1 & N–1 \\
$(m_i,\Lambda_i)$ & N–1 & N–1 \\
\bottomrule
\end{tabular}
\caption{Each cell corresponds to the value of correction factor, $k$ in different parameter subspaces for BNS \& NSBH systems.}
\label{tab:k_values}
\end{table}

Hence, for NSBH systems, the odd ratio  can be calculated as:
\begin{equation}
\boxed{
\begin{aligned}
\log O^i_j =\ & (N-1)\log f \\
& + \sum_{n=1}^N \left[\log \mathcal{Z}(H_i|d_n,I) - \log \mathcal{Z}(H_j|d_n,I)\right]
\end{aligned}
}
\label{eq:corr_odd_ratio}
\end{equation}

\subsection{Choices of Population of NSBH binaries}

\begin{figure}
    \centering
    \includegraphics[width=1\linewidth]{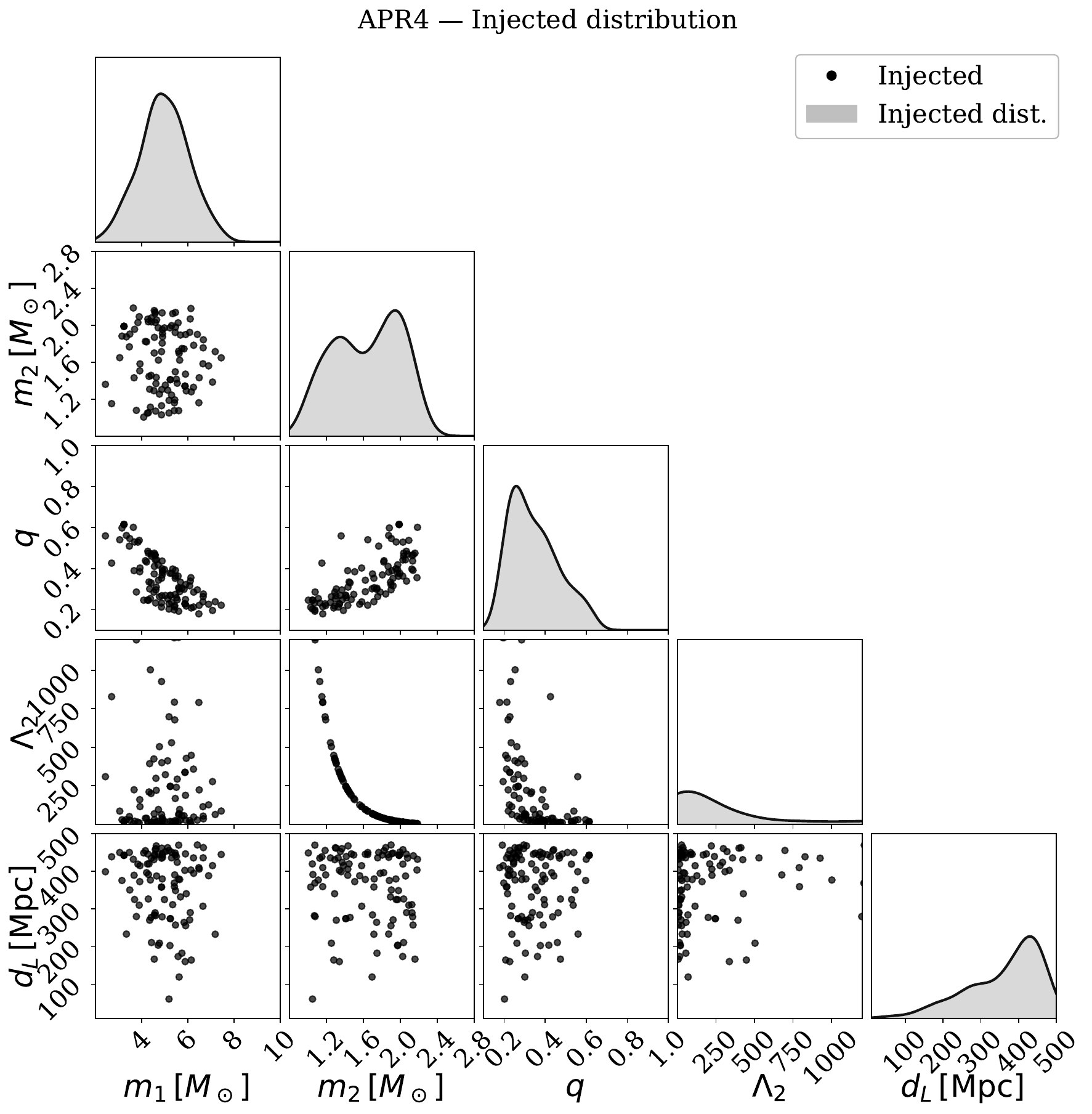}
        \caption{Injection distributions for component masses, mass ratio, tidal deformability of neutron star and the luminosity distance are shown. Although the $m_2$ distribution is uniform, the tidal deformability distribution is peaking around smaller values, because $\Lambda \propto k_2/C^5 \propto 1/m_2^5$. The higher masses have a lesser contribution to the $\Lambda$-distribution compared to the lower masses. We are not doing population constraints in this study, but only EOS constraints give certain set of observations.}
    \label{fig:APR4_injection_distribution}
\end{figure}

In this study, we accommodated the population having a Gaussian distribution in primary mass and a uniform distribution in secondary mass. In particular, $m_{BH} \in \mathcal{N}(\mu=5, \sigma=1)M_\odot$ and $m_{NS} \in \mathcal{U}(1, m^{max}_{eos})M_\odot$. Here, the lower mass gap is not taken into account. The arbitrary choice of population is taken to demonstrate the efficacy of the model selection pipeline. The luminosity distance in the comoving volume for each binary is randomly extracted from the $\mathcal{U}(10,500)Mpc$. The sky localisation parameters, i.e., $ra$, $dec$, $iota$, $psi$, are randomly taken over the whole sky. The phase in coalescence time is taken exactly 0. In FIG-\ref{fig:APR4_injection_distribution}, the injected distribution of some of the key parameters (for injected EOS APR4) is shown.

\section{Results}\label{sec:results}
\subsection{PE Results}
We performed Bayesian inference on $\sim$900 simulated NSBH events(100 for each EOS-detector pair). The simulations were carried out using with APR4, SLy \& DD2 injected EOS and $O5$, $As$ \& $ECC$ detector configurations.

\begin{figure}[h!]
    \centering
    \includegraphics[width=1\linewidth]{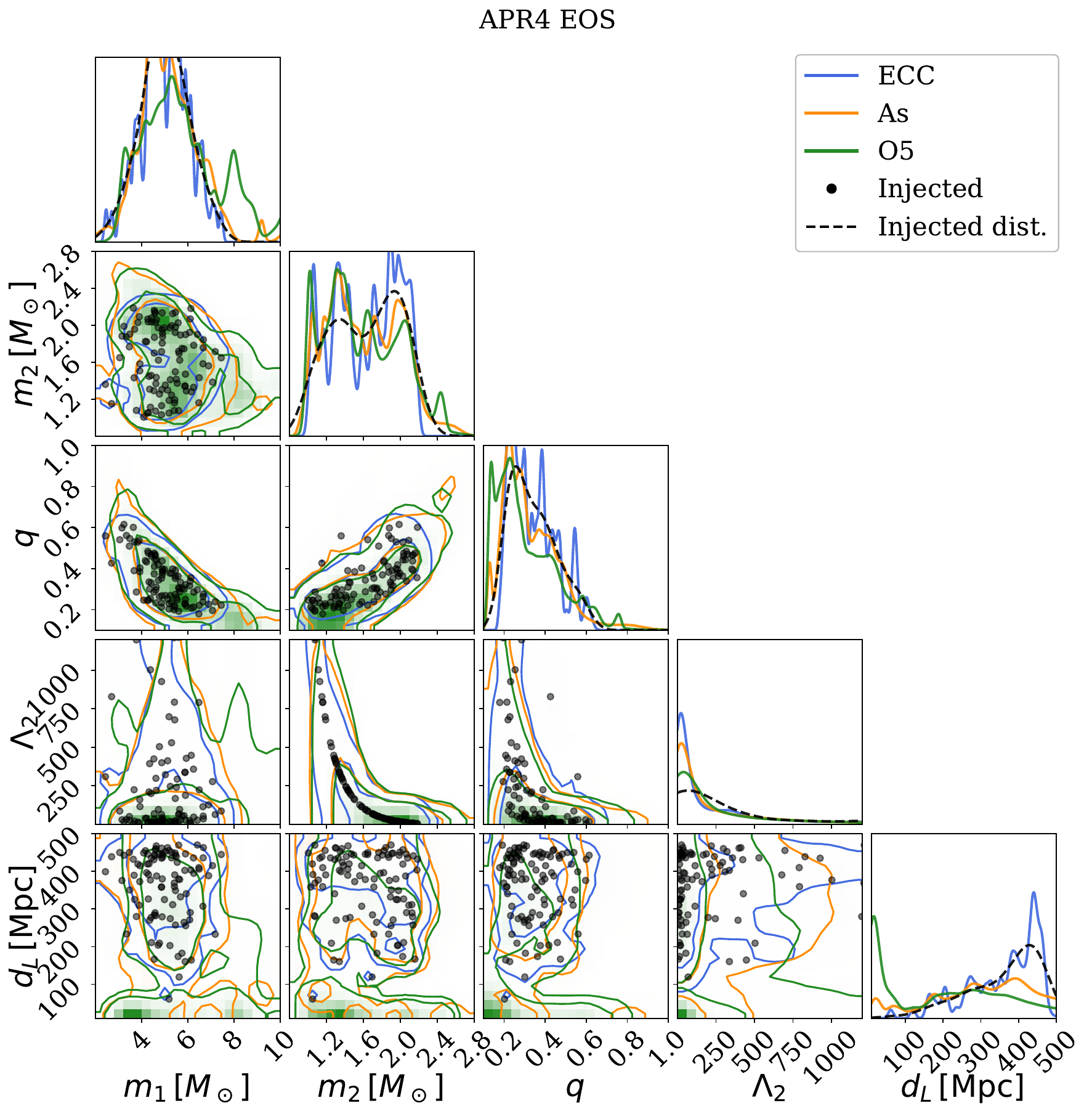}
    \caption{Comparing recovered distributions of some key parameters with their injected distribution (as shown in Fig-\ref{fig:APR4_injection_distribution}), considering \textbf{APR4} as the injected EOS. Here, the filled circles are the injected values and their probability distributions are in a black-dashed histogram in diagonal boxes. The blue, orange and green contours are the probability distributions, showing the recovered posteriors from ECC, As and O5 configurations, respectively.}
    \label{fig:corner_plot_APR4_multi_detector}
\end{figure}

In the FIG-\ref{fig:corner_plot_APR4_multi_detector}, the injected as well as recovered corner plot distribution for some key parameters are shown: $m_1$, $m_2$, $q$, $\Lambda_2$ and $d_{lum}$. The discussion of these results can be found in the upcoming sections.

\begin{figure*}[ht]
    \centering
    \includegraphics[width=1\linewidth]{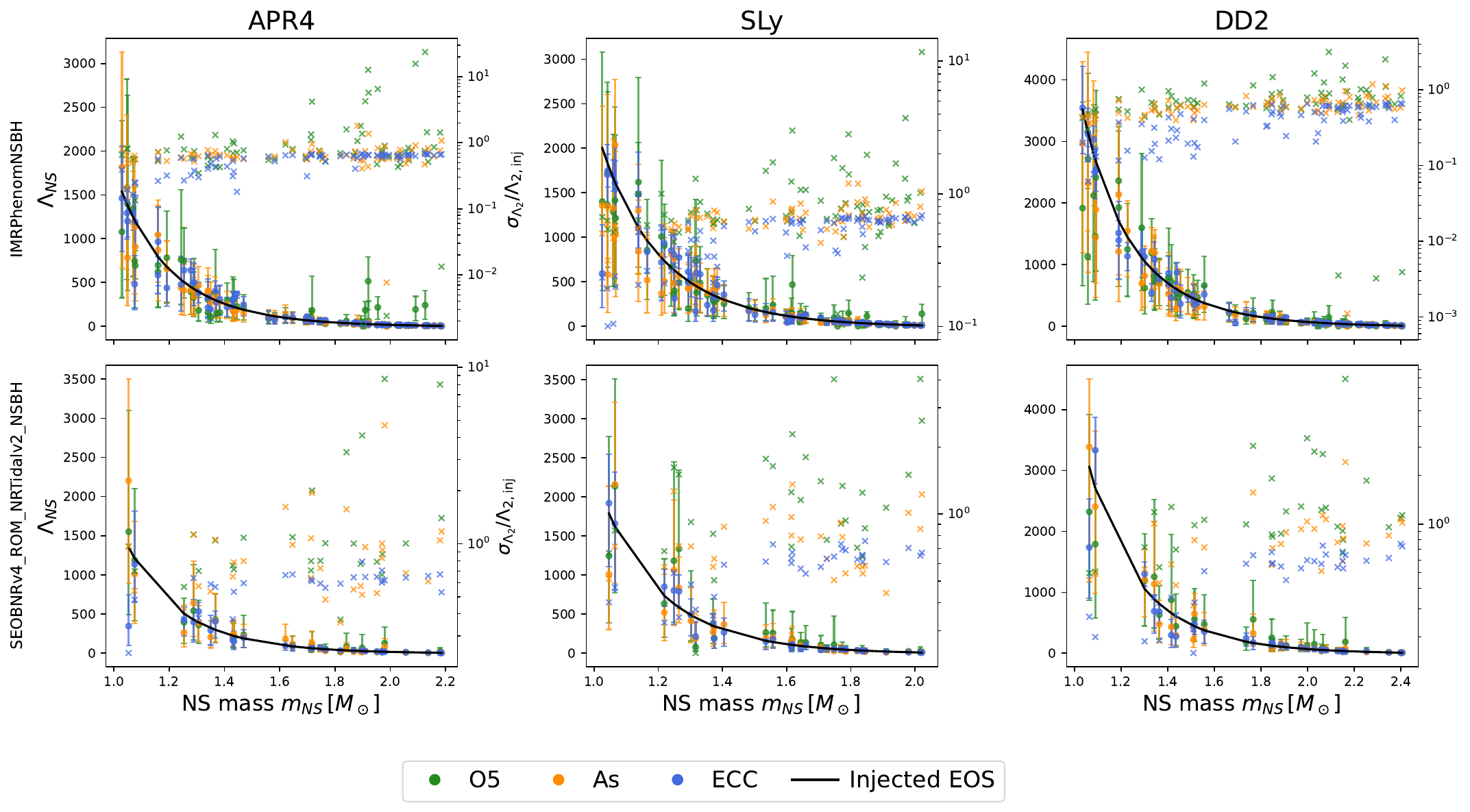}
    \caption{Recovered tidal deformability $\Lambda_2$ vs. neutron star mass $m_{NS}$ for the three injected EOS (APR4, SLy, DD2; columns) and two waveform models (rows: \texttt{IMRPhenomNSBH} (\textit{top}); \texttt{SEOBNRv4\char`_ROM\char`_NRTidalv2\char`_NSBH}, (\textit{bottom})). Circular points show the median value of the tidal deformability posterior ($\Lambda_{med}$) for each injected event with $\pm 1\sigma$ credible intervals. Black curves show the $\Lambda_2(m_{NS})$ relation for the true injected EOS (\textit{from left to right}) for APR4, ALy and DD2. Colored cross markers signifies the relative error in the recovery of tidal deformability ($\frac{\sigma_{\Lambda_2}}{\Lambda_{2,inj}}$) for O5 (green), As (orange), ECC (blue) plotted on the left axis in the subplots. Recovery is consistent with the injected relation across networks and waveform models, with larger uncertainties at higher NS mass where $\Lambda_2 \rightarrow 0$. Despite larger error at lower tidal deformability, the relative error increases for higher masses (lower tidal deformability).}
    \label{fig:m_vs_lambda_wf_comparison}
\end{figure*}

We used two NSBH-tidal waveform models: \texttt{IMRPhenomNSBH} and \texttt{SEOBNRv4\char`_ROM\char`_NRTidalv2\char`_NSBH}, to check that our EOS inference is not an artefact of waveform systematics. For each event, we recover the posterior on $\Lambda_2$ and compare it against the injected EOS relation $\Lambda_2(m_{NS})$. We repeat this for both of the waveform models, and for all three detector networks. FIG-\ref{fig:m_vs_lambda_wf_comparison} shows the recovered $\Lambda_2$ vs. neutron star mass for the 3 injected EOS, with both waveform models and all 3 detector networks overlaid. In all cases, the recovered median values with $1\sigma$ credible intervals enclose the injected $\Lambda_2(m_{NS})$ values, indicating that tidal deformability is recovered consistently regardless of which waveform model is used. The figure also shows the relative error in recovering the tidal deformability (right y-axis, shown in cross) as a function of NS mass. It is consistent with the physical expectation that with higher masses, the recovery of tidal deformability becomes more uncertain, less reliable and less able to distinguish models as NS mass increases. At the lower NS masses, different EOS predict widely varying values of $\Lambda_{NS}$. But at high masses ($\simeq 2 \ M_{\odot}$), the error in the measurement becomes of the order of the measured value itself. Consequently, the recovered values cannot effectively distinguish between different models rendering this as one of the fundamental limitation of gravitational wave observations of NS in binary systems.

\begin{figure}[h!]
    \centering
    \includegraphics[width=1\linewidth]{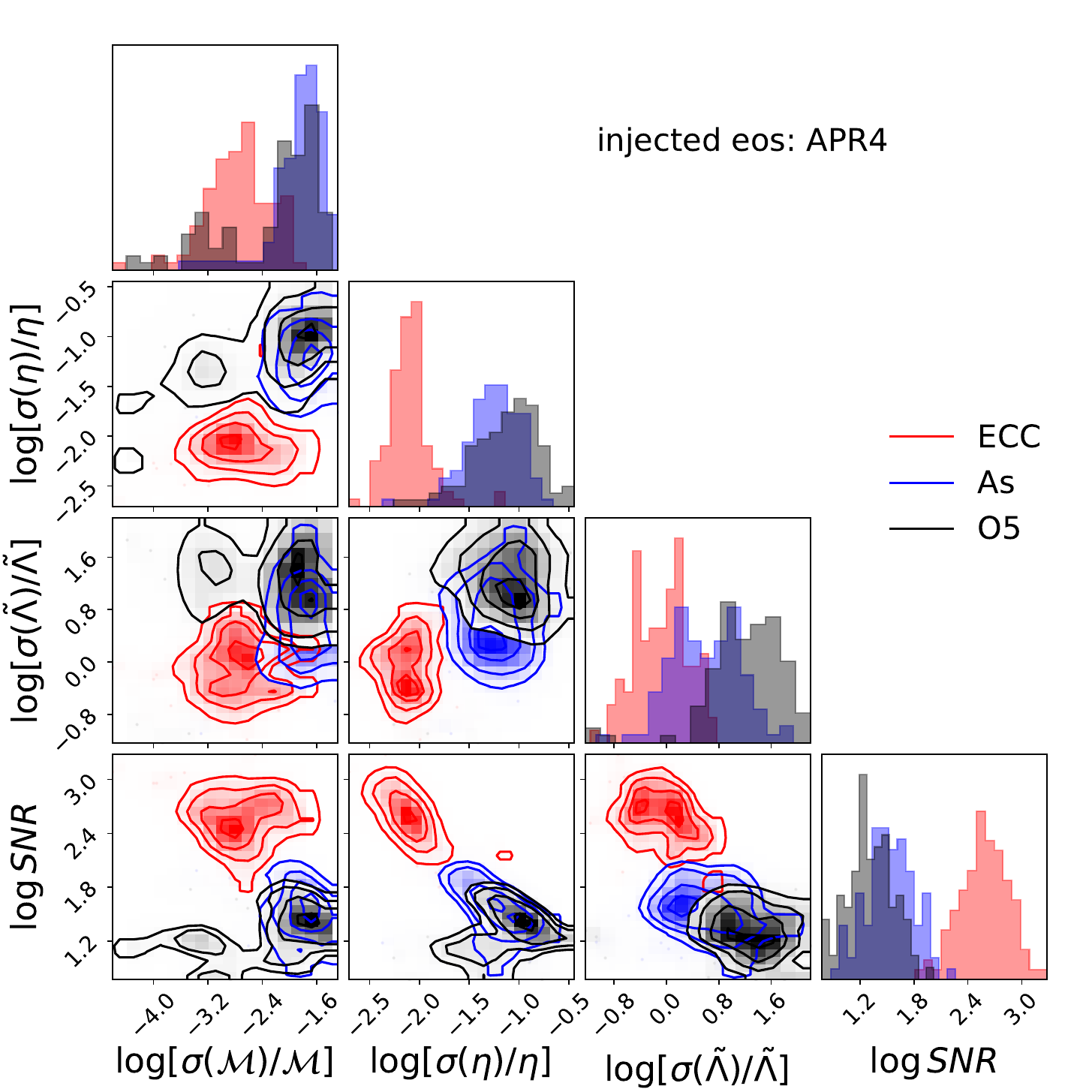}
    \caption{
    In this figure, we have shown the relative error distribution of three fundamental parameters of NSBH system $\mathcal{M}$, $\eta$ and $\tilde\Lambda$, across all $3\times100$ simulated binaries. It can be seen that $O5$ and $As$ are showing a similar kind of distribution in $\mathcal{M}$ and $\mathcal{\eta}$ errors, and for $As$, a small improvement in $\tilde\Lambda$ error distribution. For the case of $ECC$, it has a significant improvement in measuring these parameters. The $ECC$ detector also increased the sensitivity by 2 to 3 orders of magnitude in comparison to $O5$ and $As$ detectors.}
    \label{fig:Error_dist_APR4}
\end{figure}

The same parameter recovery efficiency is also plotted in FIG-\ref{fig:Error_dist_APR4}. It shows the error distributions of key parameters of binary mergers: chirp mass ($\mathcal{M}$), symmetric mass ratio ($\eta$) and effective tidal deformability ($\tilde\Lambda$). The logarithmic relative errors in $\mathcal{M}$ and $\eta$ for the O5 and As configurations are almost similar, which is further improved by the next generation ECC configuration by almost an order of magnitude. Though, tidal effects enter the phase evolution of gravitational waves at the 5th post-Newtonian (PN) order, making it very challenging to recover them precisely. Although, the $\tilde\Lambda$-distribution with ECC configuration followed the trend of error distribution similar to those of $\mathcal{M}$ and $\eta$ compared to the other configurations. In addition to these parameters, we also present the network SNR distributions for all three detector configurations. The ECC network clearly shows a 2-3 order increase in detector sensitivity compared to $O5$ and $As$. 

We compared the error distributions of our NSBH systems with those obtained for BNS systems\cite{Kashyap_2025_BNS_Bayesian_optimization}, and found that all of these parameters exhibit clearer separation in the case of NSBH systems.

\subsection{Evidence Distribution for NSBH system}
To compare evidence values for different EOS, we considered a number of tabulated model EOS from different formulation and composition classes:
\begin{itemize}
    \item Nucleonic EOS: \textbf{DD2}\cite{DD2_HEMPEL_2010, DD2_TYPEL_2010}, \textbf{SFHo}\cite{SFHo_Steiner_2013}, \textbf{APR3, APR4}\cite{APR_1998}, \textbf{SLy}\cite{SLy_Shen_2001}, \textbf{LS220}\cite{LS220_LATTIMER_1991};
    \item Hyperonic+Nucleonic EOS: \textbf{BHB}\cite{BHB_Lattimer_2014}, \textbf{H4}\cite{H4_1991, H4_2006};
    \item Incorporating phase transitions: \textbf{ALF2}\cite{ALF2_Alford_2005};
    \item Piecewise Polytropic EOS: \textbf{PP2, PP5}\cite{PP_Godzieba_2021}.
\end{itemize}

\begin{figure*}
    \centering
    \includegraphics[width=1\linewidth]{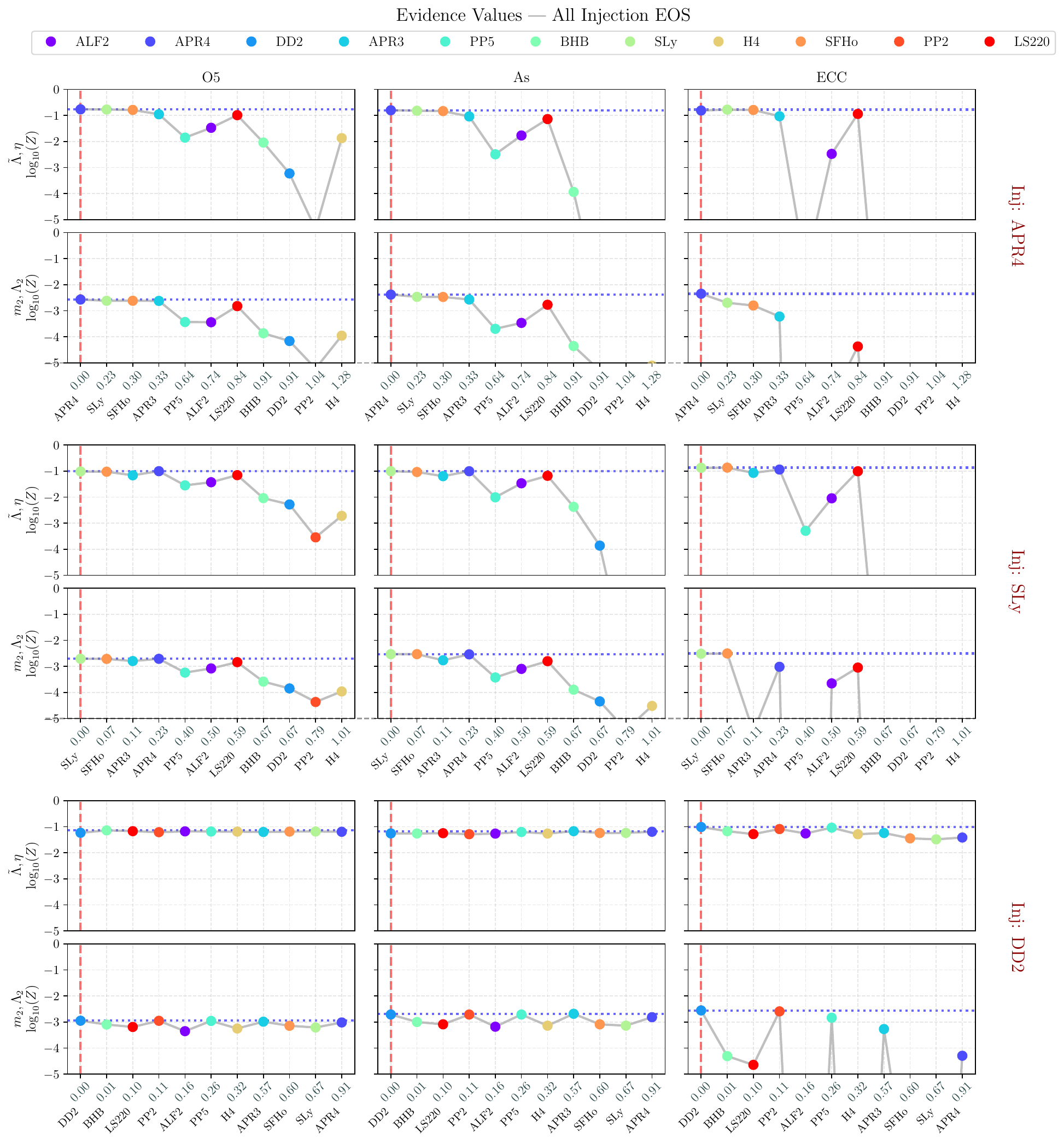}
    \caption{This figure presents the central result of our study. We have shown the median values of evidence distributions for all 100 simulated binaries in the two-parameter subspaces discussed in section-\ref{sec:evidence_calc_methods}. The first two rows correspond to APR4 as the injected EOS, followed by the SLy and DD2 for the next rows. The top panel showed the evidence values in $(\tilde\Lambda,\eta)$ space corresponding to three detector configurations, while the second panel shows evidence values in $(m_2,\Lambda_2)$ space. The model EOS on the x-axis are arranged in the order of their corresponding $L_2$-distances, which emphasize how much the model EOS differs from the injected EOS. The red dashed line is to highlight the injected EOS. With increasing sensitivity in the detector configurations, the distinguishability in the models becomes prominent. In addition to that, comparing the two subspaces, the $(m_2, \Lambda_2)$ subspace is favored over the $(\tilde\Lambda,\eta)$ subspace. Hence, for distinguishing and constraining EOS of NS in the NSBH systems, the $\boxed{(m_2, \Lambda_2) - \mathcal{ECC}}$ is the best combination.}
    \label{fig:evidence_L2_distribution}
\end{figure*}

We compute the evidence in two distinct parameter subspaces: $(\tilde\Lambda,\eta)$ and $(m_2,\Lambda_2)$ as described in sections \ref{sec:Evidence-B} and \ref{sec:Evidence-A}. The resulting evidence distributions are presented in FIG-\ref{fig:evidence_L2_distribution} for all the events. In all three detector configurations, the first 2 rows have APR4 as the injected EOS, followed by SLy and DD2, respectively. The first row shows the median evidence values in the $(\tilde\Lambda,\eta)$ subspace for all models. It is evident that for 9 of the 11 model EOS, the $log_{10}\mathcal{Z} \geq -2$, and 5 of them exhibit nearly identical maximum evidence values for O5. A similar trend persists even for the $As$ and ECC configurations. The second row, which presents the evidence values calculated in the $(m_2,\Lambda_2)$ subspace, also shows the analogous feature. However, in the ECC configuration, the distinction becomes more pronounced compared to the other two configurations, with the injected EOS (APR4) attaining the highest evidence value. This indicates that the level of distinguishability is substantially larger in the $(m_2,\Lambda_2)-ECC$ case. We tested our hypothesis using other EOS, SLy and DD2, with the consistency of the injected model yielding the highest evidence values. As expected, nearby EOS yield comparable evidence values.

FIG-\ref{fig:evidence_L2_distribution} also features the $L_2$-distances (see appendix-\ref{sec:L2_distance} for the definition of $L_2$-distance) between injected and model EOS, and arranges the models with increasing $L_2$ values. In principle, with increasing $L_2$-distance, the evidence curve should have gone downwards, but there is a mismatch between $L_2$ ordering and the evidence ordering. The main reason for this difference can be due to the differences in weighting in $L_2$ and evidence integrations. The $L_2$-distance weights all NS masses uniformly, assigning equal importance to the entire observable mass range regardless of the actual mass distribution of the detected events. Whereas, the Bayesian evidence integral is weighted by the posterior distribution of the observed events, hence it is insensitive to EOS differences outside that range and therefore, there will be no contribution of evidence in that range.

We found our results to be consistent with those obtained from BNS simulations\cite{Kashyap_2025_BNS_Bayesian_optimization}. The primary subspaces of analysis in the  study of BNS were $(\tilde\Lambda,\eta)$ and $(m_1,\Lambda_1)+(m_2,\Lambda_2)$ with an extra 4D subspace $(m_1,\Lambda_1,m_2,\Lambda_2)$. Due to the degeneracy, different $\eta-\tilde\Lambda$ pairs can yield similar kind of evidence distribution for different EOSs. Consequently, the $\mathcal{Z}_{\tilde\Lambda,\eta}$ does not provide sufficient discrimination between different EOS, making it a less effective choice for model selection for the NSBH systems as well. Instead, the $(m_2,\Lambda_2)$ subspace is preferred for model selection.

\subsection{Cumulative Evidence and Bayes Factor of EOS}\label{sec:BF_multievent}
From an underlying population of compact binary coalescence, we sample neutron star (NS) masses and their corresponding tidal deformabilities from a predefined prior distribution. For an individual NSBH event, the data provide constraints on the tidal deformability of a single NS. Similarly, a single binary neutron star (BNS) event constrains the equation of state (EOS) at only two points in the mass–deformability plane. Consequently, analyses incorporating a larger number of NSBH binaries are crucial for robust EOS model selection.

We compute and plot the cumulative Bayesian evidence for different injected EOS models as a function of the number of observed events. To reduce statistical fluctuations and mitigate selection bias, we randomly select sets of 50 simulated events and repeat this resampling procedure 20 times. As the number of events increases, the ability to discriminate between competing EOS models improves markedly: the cumulative evidence increasingly favours the true (injected) EOS, which consistently attains the highest cumulative evidence among all models considered.

Comparing the three observational configurations, we find that in the ECC configuration the cumulative evidence curve declines most steeply for incorrect EOS models. This behaviour indicates a more rapid accumulation of discriminating power, attributable to the higher signal-to-noise ratios and correspondingly tighter posterior constraints on the relevant parameters. For $O5$, both the $(\tilde\Lambda,\eta)$ and $(m_2,\Lambda_2)$ subspaces show similarly modest separation, with cumulative $log_{10}\mathcal{Z}$ spreading to roughly -1000 to -1500 over 50 events depending on the subspace. Moving to $ECC$ substantially increases the attainable separation: in the $(m_2,\Lambda_2)$ subspace, the spread grows to roughly -6000, allowing individual model EOS to be distinguished much more clearly. In contrast, the $(\tilde\Lambda,\eta)$ subspace shows comparatively weaker separation, indicating that the mass-tidal-deformability parameterization $(m_2,\Lambda_2)$ is the more discriminating subspace as sensitivity improves, while $(\tilde\Lambda,\eta)$ saturates earlier in its ability to distinguish EOS models.

\begin{figure}
    \centering
    \includegraphics[width=\linewidth]{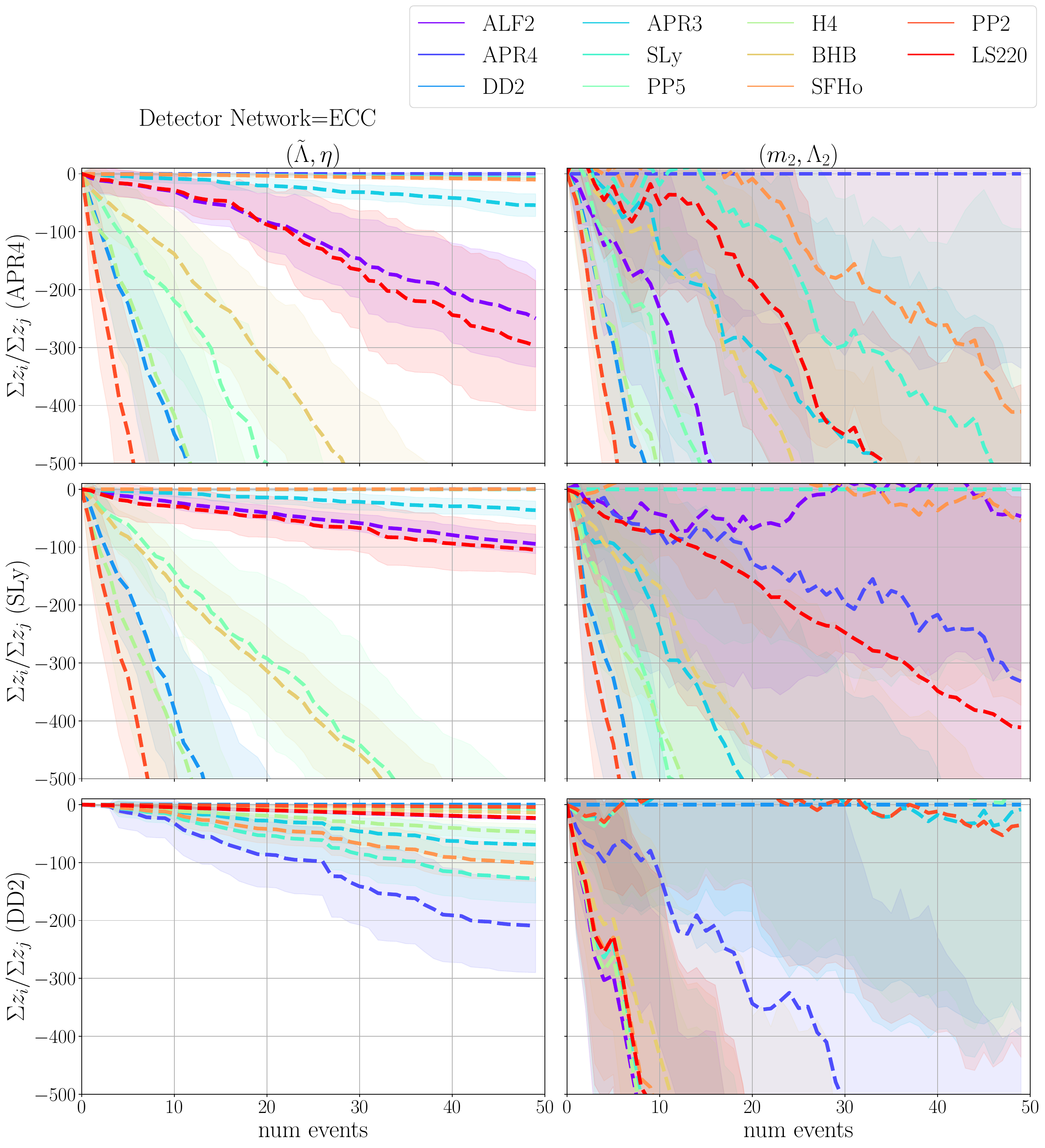}
    \caption{Cumulative log $\mathcal{BF}$ as a function of the number of NSBH events for the ECC detector network. Rows correspond to the true injected EOS (APR4, SLy, DD2, top to bottom); column correspond to the two evidence-estimation subspaces, $(\tilde\Lambda, \eta)$ (left) and $(m_2,\Lambda_2)$ (right). Each colored curve tracks one candidate model EOS from the 11-EOS set; shaded bands show the $1\sigma$ spread over bootstrap resamplings of the event catalog. Curves that fall away from zero indicate that the corresponding model EOS is increasingly disfavored relative to the true injected EOS as more events are combined.}
    \label{fig:BF_ECC}
\end{figure}

The cumulative evidence can tell us which model EOS is statistically preferred over others and what the order of preference will be from a given list of EOS. But to quantify this relative model preference more explicitly, we next computed the Bayes Factor $(\mathcal{BF})$ as a function of the number of events, which allows us to assess the statistical confidence in favouring one EOS over another. FIG-\ref{fig:BF_ECC} shows the Bayes' Factor as a function of the number of events for the ECC detector configuration. The rows correspond to the three injected EOS in the order of increasing stiffness from top to bottom. As the number of events increases, the $\mathcal{BF}$ trends towards stabilisation, indicating convergence in model preference. For the ECC configuration, this convergence occurs significantly faster compared to O5 and As due to its superior sensitivity. However, these curves are not strictly monotonic, reflecting event-to-event variance. A small number of statistically bad events can temporarily flatten or reverse the trend before the larger sample size recaptures the dominant falling trend. Furthermore, the $L_2$ distance does not capture the nuanced differences between two EOS. For example, two EOSs with intersecting $m-\Lambda$ curves will have similar evidence if the NS masses in the NSBH population are clustered near the intersection region. Given these minor caveats, the principal conclusion of our work is to demonstrates that given a sufficiently large NSBH population ($\sim 20-30$) and advanced detector sensitivity, Bayesian model selection can robustly identify the underlying neutron star EOS.

\subsection{Evidence of LVK NSBH events: GW200105, GW200115, GW230529}
We calculated the evidence in the parameter space discussed earlier for the real events reported by LIGO-Virgo-KAGRA (LVK) collaboration. The GW200105 and GW200115 \citep{Abbott_2021_NSBH_discovery} events detected in O3, while GW230529\citep{Abac_2024_GW230529} event was detected in O4a observing cycle. We used the publicly available data of posterior samples of all 3 events from the Gravitational Wave Open Science Centre (GWOSC).

\begin{figure}
    \centering
    \includegraphics[width=\linewidth]{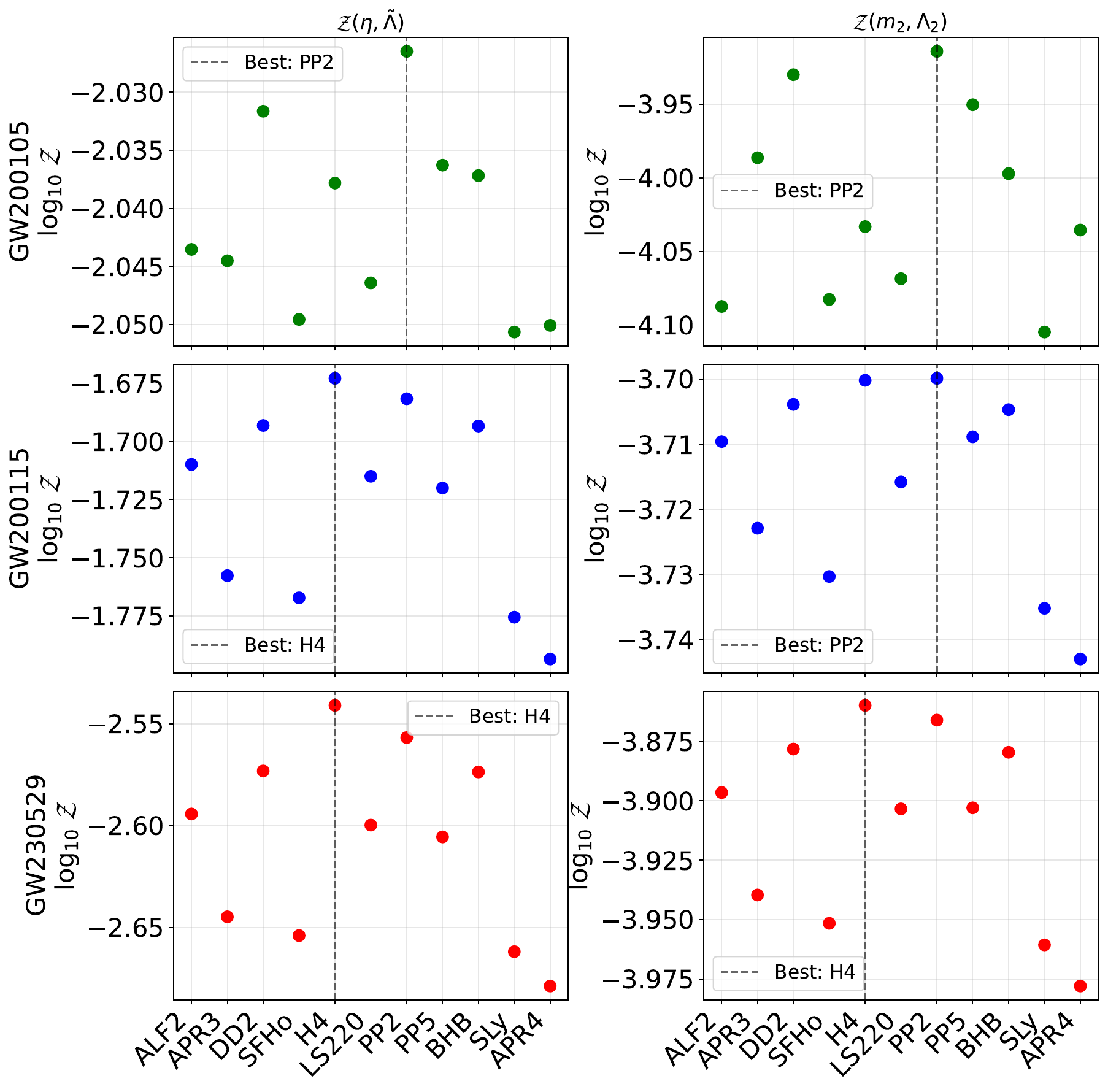}
    \caption{Evidence values in two parameter spaces: $(m_2,\Lambda_2)$ - \textit{left}, and $(\eta,\tilde\Lambda)$ - right for 3 events: GW200105, GW200115 and GW230529. The vertical dashed lines show the EOS corresponding to the maximum evidence.}
    \label{fig:LVK_event_evidence}
\end{figure}

Figure-\ref{fig:LVK_event_evidence} shows the evidence values of those events in the two parameter subspaces, namely i.e.; $(\eta,\tilde\Lambda)$ and $(m_2,\Lambda_2)$, using the calculations discussed in the section-\ref{sec:evidence_calc_methods}. The analysis showed that for GW200105 and GW200115, PP2 is preferred over other EOS while for GW230529, the data showed the preference for H4. For actual observed events, the only feasible form of discrimination is to rank the EOSs according to their relative evidence values rather than computing Bayes factor against a "true" EoS, which is not known a priori. Moreover, the uncertainties in $m-\Lambda$ posterior with current-generation detectors yield comparable evidence for a broad class of EoSs making it impossible to draw definitive conclusion for any specific EOS. 

\section{Discussion and Conclusion}\label{sec:discussion}
We present constraints on the neutron-star equation of state (EOS) inferred from a simulated population of neutron star-black hole (NSBH) binary mergers, focusing on capabilities of the current generation of ground-based gravitational-wave detectors, their planned upgrades and prospective next generation ground-based observatories in both the United States and Europe. Our analysis employs full 17-dimensional Bayesian parameter estimation for each event. We find that the median values of the posteriors reproduce the injected population approximately for masses and tidal deformabilities as well as distances in our most advanced detector configuration, ECC. From the resulting posterior distributions for the component masses and tidal deformabilities, we compute Bayes factors to assess the relative support for different candidate EOS models. This analysis framework directly connects measurement precision (set by detector sensitivity, orientation, and distance) to EOS distinguishability through the tidal imprint on the late inspiral phasing. A population-informed EOS inference is left to our future work.

Within this framework, we explore two different parameter spaces in which we performed the evidence calculation. One of the central result of the present work is that using the $(m_{NS},\Lambda_{NS})$ space yields stronger discriminatory power (requiring about less number of events, $\sim 20$) between EOS models than the $(\tilde{\Lambda},\eta)$ space. We attribute this improvement to parameter degeneracies involving $\tilde{\Lambda}$ and $\eta$ with individual component masses and tidal deformabilities, which obscure the distinctions between different EOS models, while working directly in $(m_{NS},\Lambda_{NS})$ mitigates these degeneracies and leads to higher evidence for the EOS present in the data. We applied our pipeline to three NSBH events reported by the LVK collaboration using publicly available data. Within the limitations of SNRs imposed by current SNRs, we find a largest support for PP2 and H4 models however, the evidence is not convincing with so few events. Constraining equation of state of neutron star and distinguishing the binaries containing neutron star are deeply interconnected goals in gravitational wave and multimessenger astronomy\cite{Bhaskar_2025_constraining_NS_EOS}, which we plan to undertake in our future work.  These results highlight the importance of detector sensitivity and analysis choices for robust EOS inference, and motivate future extensions incorporating hierarchical population inference (including selection effects), broader EOS parameterisations, and improved waveform modeling for precession and higher-order modes.

\begin{acknowledgments}
This work was carried out as part of VD's master's thesis project at IIT Bombay. We are grateful to Archana Pai and the Astrophysics, Cosmology and Gravity (ACG) group at IIT Bombay. We thank Ish Gupta for many discussions, insightful feedback, and assistance in incorporating the PSDs of future detectors into the Bilby pipeline. We also thank Bangalore Sathyaprakash and the GW group at Penn State, where the initial phase of this work was completed. This material is based upon work supported by NSF's LIGO Laboratory which is a major facility fully funded by the National Science Foundation. RK and VD acknowledge the support of Param Rudra, the high‑performance computing facility established under the National Supercomputing Mission at IIT Bombay. We acknowledge the the LVK computing cluster and LISC, as well as the various Python packages that supported our research.
\end{acknowledgments}

\section*{Contributions}
VD has performed the Bayesian inference runs and written scripts for analysing the runs, including the public GW data of LVK events. VD has written major part of the paper. RK has conceptualised the problem, organised writing and advised VD as part of his M.Sc. thesis on this work. YB had done initial Bayesian runs and has contributed to preparing analysis scripts and writing the introduction. AI has been used for grammatical check, rephrasing about 20\% of the sentences and preparing parts of Bayesian inference analysis scripts.

\bibliography{NSBH-Mergers, NSBH-Mergers_url}

\newpage
\appendix \label{sec:appendix}

\section{Computational Cost Optimization}\label{sec:cost_effective}
We follow \citet{Veitch_Hu_2024_Cost_of_Bayesian_PE} to examine and optimize the computational challenges faced by standard Bayesian Parameter Estimation and its comparison with some accelerated methods. With next-generation ground-based detectors, the detection rate and SNR will be higher, the standard methods will demand up to a quadrillion CPU hours every month. The accelerated methods like relative binning\cite{relative_binning_method_paper_Zackay_Dai_Venumadhav_2018} (RB), and relative order quadrature (ROQ) will still require about million CPU hours per month for faithful PE. These methods make the estimation better, but are still very far from the ideal, as this amount of CPU hours will be a substantial burden on the computing infrastructure, electricity cost and environment \cite{Veitch_Hu_2024_Cost_of_Bayesian_PE}.

\begin{figure}[ht]
    \centering
    \includegraphics[width=1\linewidth]{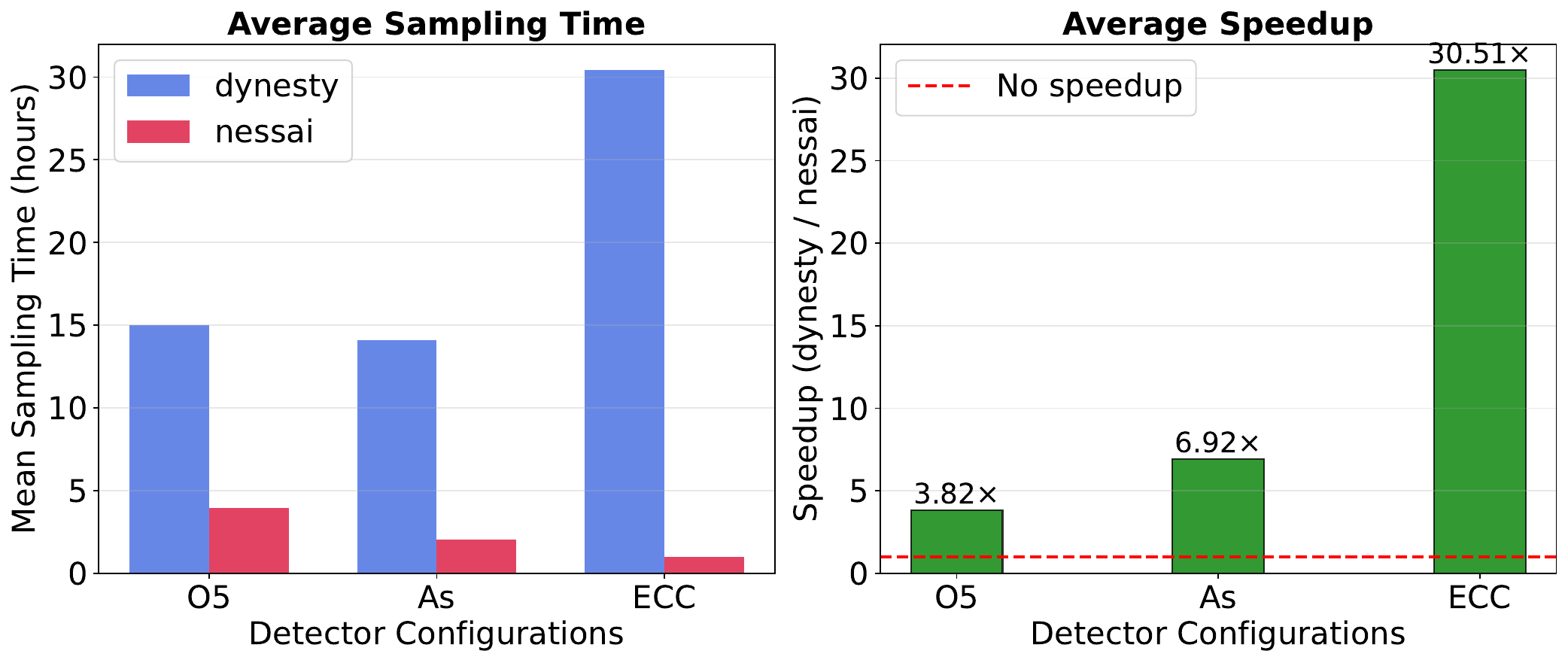}
    \caption{\textit{(left)} The average sampling time for each detector configuration are shown. It can be observed that with \texttt{dynesty}, ECC runs took a long time to complete. \textit{(right)} The comparison between the average sampling time is shown here. With the $O5$-configuration, the sampling time was reduced by more than a factor of 3 (at least). For As, the speedup is $>6$. For the $ECC$-configuration, this rises to more than $30\times$ factor.}
    \label{fig:speedup}
\end{figure}

In addition to using these different methods in the frequency domain source model and likelihood calculation, the usage of different sampling can also be taken into account. For our population study and evidence calculation, nested sampling is more effective in comparison to the Monte Carlo-Markov Chain (mcmc) sampling, as nested sampling simultaneously calculates the Bayesian evidence which is important for model selection. We incorporated two samplers, namely \texttt{dynesty}\cite{dynesty} and \texttt{nessai}\cite{nessai,Williams:2021qyt,Williams:2023ppp} for this purpose. We investigated the comparison between these samplers in the context of speed-up and log-evidence calculations. Indeed, \texttt{nessai} outperformed \texttt{dynesty} by a (mean) factor of $\sim 3.5$ in the $O5$-configuration, which improved by a lot in $ECC$-configuration by $\sim 31 \times$. See FIG-\ref{fig:speedup} for further details.

\section{$L_2$ distance between model EOS}\label{sec:L2_distance}
\begin{figure}
    \centering
    \includegraphics[width=\linewidth]{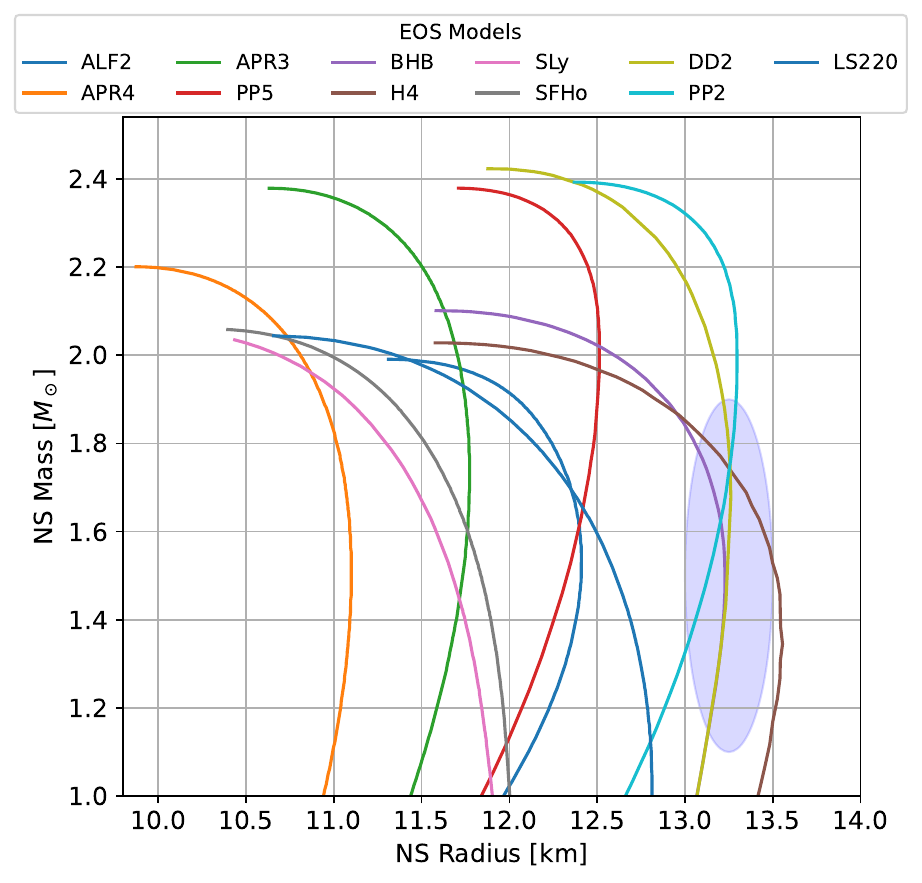}
    \caption{Mass vs Radius curves of all the model EOS considered. The shaded elliptical part is showing the difference between two nearby EOS: DD2 and BHB.}
    \label{fig:EOS_MvsR}
\end{figure}

The $L_2$ distance is used as a metric to quantify the disparity between a pair of model EOS of neutron stars in terms of their mass-tidal deformability ($m-\Lambda$) curves\cite{Kashyap_2025_BNS_Bayesian_optimization}. The $L_2$-distance between the mass-lambda curves of two model EOSs, A and B, is defined as:
\begin{equation}
\boxed{
    L_{2,\Lambda}^{A,B} \equiv \frac{\int_{m_l}^{m_u}[\Lambda_A(m)-\Lambda_B(m)]^2 dm}{\sqrt{\int_{m_l}^{m_u}\Lambda_A^2(m)dm \times\int_{m_l}^{m_u}\Lambda_B^2(m)dm}}
    }
\end{equation}

where $\Lambda_A(m)$ and $\Lambda_B(m)$ are the mass-lambda curves for the models A and B (see, for example, DD2 and BHB in FIG-\ref{fig:EOS_MvsR}). The $m_l$ is the smallest NS mass in the observed population taken to be 1 M$\odot$, and $m_u$ is the smallest of two maximum allowed masses corresponding to the two models.

\end{document}